# All-Lossy Quasi-Guided Dual-Mode Optical Waveguide Exhibiting Exceptional Singularities

Arnab Laha,[1] Abhijit Biswas,[2] and Somnath Ghosh [1, *]

[1] *Department of Physics, Indian Institute of Technology Jodhpur, Rajasthan-342037, India*
[2] *Institute of Radio Physics and Electronics, University of Calcutta, Kolkata-700009, India*
*Corresponding author: somiit@rediffmail.com*

**Abstract:** We explore exceptional points (*EP*) in a dual-mode symmetric planar optical waveguide with transverse variation of inhomogeneous loss profile; where modal evolution alongside an *EP* is reported in the context of selective optical mode conversion.

## 1. Introduction

Lately, the quantum inspired analogous dissipative optical systems enriched with enticing characteristics of non-Hermitian quantum mechanics have been attracted enormous attention in the field of photonics research. In this context, occurrence of exceptional points (*EP*) in such non-Hermitian optical systems is one of the interesting aspect in recent years. Essentially *EP*s refer to the branch point singularities in $2D$ system parameter space where two coupled eigenstates coalesce. The topological characteristics of an *EP* can be tuned through system parameters viz. geometrical parameters, non-hermitian effects in terms gain and/or loss, etc. An adiabatic variation of two coupling parameters along a closed loop around the *EP* allows the continuous swapping between the coupled states quasi-statically [1-4]. This intriguing phenomenon alongside an *EP* opens a possibility of selective mode conversion between the coupled modes i.e., in principal switch either from lower to higher order mode or vice-versa. Such fascinating physical effects associated with appearance of an EP have been recently explored in various open optical systems like optical waveguides [1,2], partially pumped optical microcavities [3] and laser systems [4], etc.

In this paper, we report a $1D$ symmetric planar dual-mode optical waveguide to explore controllable optical mode conversion around an EP. Here non-Hermiticity is attained by injecting transverse inhomogeneous loss profile only i.e. avoiding the $\mathcal{PT}$-symmetric constraints; where an *EP* is encountered by tuning the loss level (in terms of a tunable co-efficient $\gamma$) and a fractional loss difference (dictated by a tunable factor $\tau$) in the core region. Contextually, several waveguide geometries have been reported with presence of both gain and loss distribution [1]. However, according to the proposed design no external gain is required during operation; where one may able to control the coupling between the leaky modes by maintaining the inhomogeneity in attenuation-profile only. Contextually, such a specific no-gain all lossy optical structure has been reported for the first time to the best-of-our-knowledge. Encountering an *EP* in the controlling parameter space of the specially configured optical waveguide, we analyze the modal dynamics in terms of evolution of two supported modes with simultaneous study of the characteristics of corresponding propagation constants around the *EP* towards selective optical mode switching.

## 2. Optical waveguide configuration and modal characteristics

Accordingly, we consider a $1D$ symmetric step-index planar optical waveguide, with suitably customized transverse non-uniform loss profile. For a steady-state transverse mode $\psi(x)$ with frequency $\omega$ and propagation constant $\beta$, the modal equation can be written as $[\partial_x^2 + \omega^2 n^2(x) - \beta^2]\psi(x) = 0$ (considering the propagation along the $z$-axis w.r.t. transverse $x$-axis). Under operating condition, the schematic of the proposed geometry of the waveguide with transverse distribution of complex refractive index profile $n(x)$ and relative permittivity profile $\epsilon(x)$ are shown in Fig. 1(a). The passive refractive indices of core and cladding regions are chosen as $n_l (= 1.46)$ and $n_h (= 1.50)$ respectively. Such prototype of the waveguide structure can be fabricated with thin-film deposition of glass materials over a thick silica glass substrate. The total width of the waveguide is set at $W = 40$ in dimensionless unit (normalizing $\omega = 1$; where $W = 20\lambda/\pi$ with free-space wavelength $\lambda$). All the parameters are chosen in such a way that only two modes i.e., the fundamental mode ($\psi_{FM}$) and the first-higher-order mode ($\psi_{HOM}$) are supported by the waveguide.

During operation, the overall transverse attenuation level is tuned by the loss-coefficient $\gamma$ (starting from 0 up to 0.5), where inhomogeneity in the core imaginary index profile is maintained by the fractional parameter $\tau$. With introduction of inhomogeneous loss in the waveguide, two leaky modes are mutually coupled and corresponding propagation constants ($\beta$) exhibit avoided resonance crossing (ARC) with crossing/ anti-crossing of their real and imaginary parts. The abrupt change in two dissimilar behavior of ARCs as shown in Fig. 1(b) for $\tau = 1.165$ (upper panel) and $\tau = 1.175$ (lower panel) confirms the presence of an EP [1,3] in system parameter space at ~ ($\gamma_{EP}$ = 0.2008, $\tau_{EP} = 1.17$). To study the unconventional physical effects around the identified EP, we enclose it in ($\gamma, \tau$)-plane as shown in Fig. 1(c); where two different enclosing contours (blue and brown loops) around EP is considered to study the dynamics of $\beta$-values corresponding to the pair of coupled modes with simultaneous evolutions of associated eigenfunctions near EP. Now one anticlockwise round along the blue circular contour in ($\gamma, \tau$)-plane causes the permutation between two interacting modes (i.e. exchange in positions) in complex $\beta$-plane as depicted in Fig. 1(d), exhibiting EP as the 2$^{nd}$ order branch-point for eigenvalues. Now, the evolutions of underlaying eigenfunctions $\psi_{FM}$ and $\psi_{HOM}$ are also studied in Fig. 1(e) & (f) respectively via plotting the modal intensity profiles $|\psi(x)|^2$ for each value of $\phi$ along the brown contour (as shown in Fig. 1(c)). As can be seen in Fig. 1(e) & (f) both the eigenmodes exchange their identities among each other near and around EP.

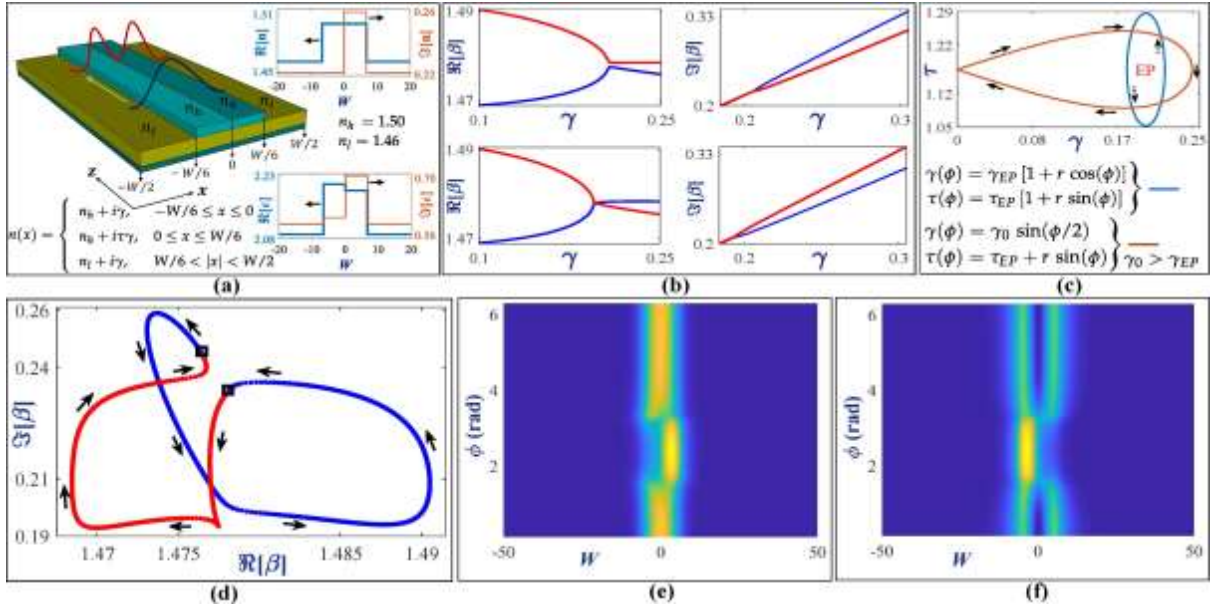

**Fig. 1: (Color online) (a)** Schematic of the waveguide structures with transverse refractive index and relative permittivity profiles; showing complex variations under operating conditions at EP. **(b)** Variations of complex propagation constants ($\beta$) w.r.t. $\gamma$; showing anticrossing in $\Re(\beta)$ and crossing in $\Im(\beta)$ for $\tau = 1.165$ (upper panel), while vice-versa for $\tau = 1.175$ (lower panel) respectively. **(c)** Enclosing an EP in parameter space described by two different sets of parametric Eqns. **(d)** Complex trajectories of $\beta$ for two coupled modes exhibiting mode flipping in complex plane (red dots for $\psi_{FM}$ and blue dots for $\psi_{HOM}$) followed by circular variation of the coupling parameters in anticlockwise direction (blue variation in Fig. (c)) around EP in ($\gamma, \tau$)-plane. Evolution of **(e)** $\psi_{FM}$ and **(f)** $\psi_{HOM}$ around the EP (mode stacking) following the progressions of $\gamma$ and $\tau$ along the loop is described by brown variation in Fig. (c) in clockwise direction.

## 4. Summary

In summary, an explicit configuration of a non-$\mathcal{PT}$-symmetric step-index planar dual-mode optical waveguide with suitably-tailored transverse inhomogeneous loss profile in the optical medium is reported for the first time; where an EP is encountered by tuning only the inhomogeneity in the loss level. Modal evolutions with simultaneous characteristics of the corresponding propagation constants ($\beta$) are investigated around the identified EP. The coupled eigenstates exhibit mode switching selectively in complex $\beta$-plane followed by an adiabatic variation of $\gamma$ and $\tau$ along a closed loop around the EP; where also the corresponding eigenmodes exchange their identities near EP. Such rich physical aspects with specific relationship between mode conversions with contour parameters near EP explore an extensive platform to fabricate on-chip state-of-the-art integrated photonic devices.

The work is partially funded by Department of Science and Technology (DST), India [IFA-12, PH-23].